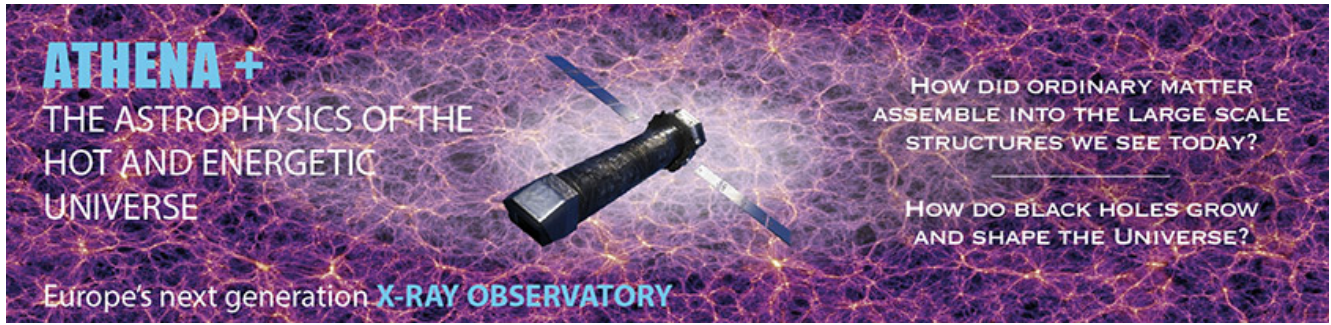

# The Hot and Energetic Universe

An *Athena+* supporting paper

## Star formation and evolution

Authors and contributors

**S. Sciortino, G. Rauw,** M. Audard, C. Argiroffi, Y.-H. Chu, M. De Becker, J. Drake, E. Feigelson, E. Gosset, N. Grosso, M. Güdel, M. Guerrero, A. Hervé, J. Kastner, R. Montez, Y. Nazé, L. Oskinova, B. Stelzer, A. ud-Doula

## 1. SUMMARY

*Athena+* will allow us to probe high-energy phenomena occurring at all evolutionary stages of stars, over almost the entire range of stellar masses, from their formation to their death. These phenomena have profound impact in star and planetary system formation as well as in feedback processes on Galactic scales.

Time-resolved high-resolution spectroscopy of bright Classical T Tauri Stars down to 3 ks with *Athena+* X-IFU will allow us to explore the dynamics of the accretion process and to study the excitation mechanism of the Fe Kα 6.4 keV line emitted from the circumstellar disk. The simultaneous observation of many density-sensitive triplets enables us to investigate the controversial issue of density stratification of accreting material. In young active stars, *Athena+* X-IFU will allow us to probe the initial phase of the most extreme flares as well as their perturbing influence on circumstellar disks of Young Stellar Objects. At the lower mass end of the Hertzsprung-Russell diagram, *Athena+* observations will tell us to what extent dynamos operating in Ultra-Cool Dwarfs produce solar-like coronal magnetic fields, or whether they generate planet-like magnetospheres with completely different physics of magnetic heating, wind mass loss, etc. For the most massive stars, high-resolution X-ray spectroscopy provides a powerful probe of their stellar winds, the degree of wind inhomogeneity and the shape of wind structures. *Athena+* X-IFU will significantly enlarge the sample of high-resolution spectra of massive stars, and open up the field of time-resolved high-resolution spectroscopy. This will enable us to study, for the first time, the dynamics of the hot plasma in the innermost parts of magnetospheres of massive stars, as well as in the inner regions of wind interaction zones of massive binaries. Finally, X-IFU spectra of the faint, soft, diffuse X-ray emission from Planetary Nebulae will enable the first systematic study of the properties of the hot bubbles inside these nebulae, revealing the origin of this hot plasma.

## 2. X-RAY EMISSION FROM YOUNG STELLAR OBJECTS

Apart from the very early protostars, the so-called Class 0 sources, in which the occurrence of X-ray emission is still highly controversial, more evolved Young Stellar Objects (YSOs) – the so-called Class I-III sources – are strong X-ray sources. Understanding their X-ray emission is essential since it deeply affects the chemistry of the circumstellar disks, and ionizes and heats circumstellar material. Disk heating provokes disk photo-evaporation and therefore limits planet formation processes or planet migration. X-ray disk ionization (several orders of magnitude larger than ionization by cosmic rays in the inner disk, e.g. Glassgold et al. 2000) couples the gas to the magnetic fields via the magneto-rotational instability, inducing MHD turbulence, and affecting angular momentum transport. Those in turn affect the protoplanetary disk evolution and, eventually, the planetary system formation and evolution. Both ionization and heating of disk material also fundamentally influence the chemical processing of disk material (Glassgold et al. 2012), and therefore contribute to the formation of molecules indispensable for the formation of habitable planetary environments (e.g., $H_2O$, perhaps also pre-biotic molecules).

**Accretion Phenomena in YSOs:** Only recently has it been found that soft X-ray emission (E ≲ 1 keV) in some YSOs is likely due to accretion shocks (e.g. Kastner et al. 2002; Argiroffi et al. 2007). In those systems material falls onto the central star with velocities ~300–500 km s$^{-1}$, forming a shock near the stellar surface. This shock heats the infalling material up to temperatures of a few MK, producing an X-ray bright "spot", as proposed by Gullbring (1994). X-ray luminosities of the shock-heated plasma are predicted to be high even for moderate accretion rates (i.e., $L_X$ ~ $10^{30}$ erg s$^{-1}$ for dM/dt = $10^{-10}$ $M_{sun}$ yr$^{-1}$, Sacco et al. 2008). A key diagnostic to distinguish between coronal and accretion shock X-rays are densities: this requires high-resolution X-ray spectroscopy. With Chandra and XMM-Newton gratings the analysis of the He-like line triplets (cf. Porquet et al. 2010) has been possible only for the 7 brightest and nearest Classical T Tauri Stars (CTTSs) and has yielded densities significantly higher than that of coronal plasma, thus in agreement with model predictions for accretion shocks. Hence in those CTTSs the soft X-ray emission provides a direct probe of the accretion process. However even for this handful of sources the statistics of counts accumulated in the triplet lines is poor and has prevented, so far, the derivation of possible Doppler line shifts due to the bulk motion (at few hundreds km s$^{-1}$) of accreting material. The current evidence shows that Class I protostars rotate faster than more evolved YSOs but well below the break-up velocity (cf. White et al. 2007); however in the case a very fast rotating YSOs (~350 sin *i* km s$^{-1}$ for $P_{rot}$ ~ 1 day and 5 $R_{sun}$ radius) exists, such a rotation would produce a Doppler shift detectable and measurable with the X-IFU in the Fe 6.7 keV line emitted by the coronal plasma.

Thanks to the *Athena+* X-IFU we can

*i)* obtain good S/N high-resolution spectra, in reasonable exposure times (≤ 50 ks), for CTTSs up to 500 pc (cf. Fig. 1, right). This will allow us to enlarge the sample size and extend these studies to all major nearby star forming regions (i.e. Taurus, Chameleon, Orion etc.);

*ii)* obtain good S/N high-resolution spectra for the brightest and nearest CTTSs on time scales of just a few ks (cf. Fig. 1, left). This will open up the possibility to look for accretion events directly from X-ray lines (Dupree et al. 2012), to explore the variability of the accretion process (on predicted time-scales ≤ 1 day according to the variability found in the optical), as well as the rotational modulation due to accretion stream shadowing. In the case of favourable orientation and relatively moderate absorption, the X-IFU will allow us to probe the stream bulk velocity;

*iii)* measure simultaneously with good S/N the line fluxes from several He-like ions (O VII, Ne IX, Mg XI), as required to probe regions of different densities and to test the scenario of accretion-fed coronae that was proposed as an alternative explanation of the soft X-ray excess observed in CTTSs (Brickhouse et al. 2010). Moreover it will be possible to measure the line fluxes of the Lyman series of OVII and OVIII to derive simultaneously the source optical depth (which is predicted and has been observed, Argiroffi et al. 2009) and the absorption due to surrounding material.

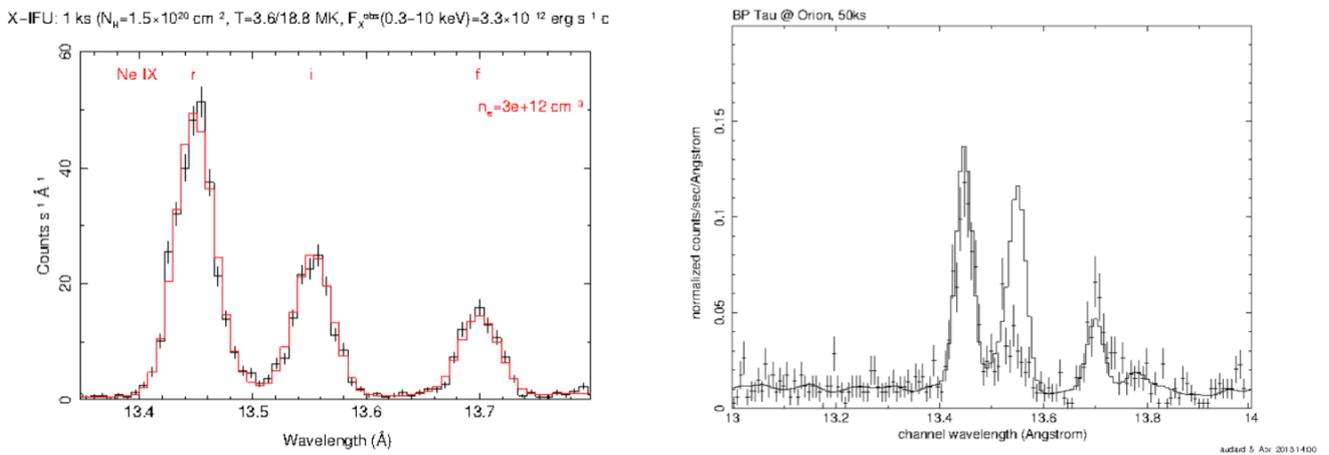

Figure 1: Predicted X-IFU spectra of the Ne IX triplet emitted from a high density (ne =3 $10^{12}$ cm$^{-3}$) plasma with a temperature of 3.1 MK: [Left] For TW Hya, the nearest (d ~ 50 pc) and X-ray brightest (fX ~ 5 $10^{-12}$ erg cm$^{-2}$s$^{-1}$, AV~0) CTTS, the spectrum can be collected with an exposure time as short as 1 ks, allowing, for the first time, a detailed analysis of the variability of the accretion process in the X-ray bandpass. [Right] For a CTTS with properties similar to those of BP Tau: fX ~ 3 $10^{-13}$ erg cm$^{-2}$s$^{-1}$, AV ~1.5, but at the distance of Orion (500 pc). In this case the integration time required for such a high-quality spectrum increases to 50 ks (it would be just ~ 3 ks at the Taurus distance). The model histogram shows the predicted triplet shapes for high-density coronal plasma as observed in BP Tau, while the data points have been computed for low-density coronal plasma.

**Probing the YSO circumstellar disks with the Fe K alpha 6.4 keV fluorescent line:** Fe K alpha 6.4 keV emission has now been observed from about a dozen YSOs (e.g. Tsujimoto et al. 2005). The presence of accretion disks together with geometrical arguments and the high value of the line Equivalent Width (EW) led to the conclusion that the Fe K alpha 6.4 keV lines from YSOs are produced in the circumstellar disk and not in the photospheres of the central stars. Today we can detect this line only in CCD-resolution spectra where often the 6.4 keV feature is barely separated from the coronal 6.7 keV line. In the small sample of Fe K alpha 6.4 keV detections surprising discoveries have been made.

The rare cases where time-resolved spectroscopy has been obtained showed that the fluorescence due to the thermal photo-ionization scenario may not always apply. Indeed, *i)* an exceptionally large equivalent width of 1400 eV was measured in the rise phase of a flare of a YSO in Orion (Czesla & Schmitt 2007). This can be reconciled with a photo-ionization origin only by making the ad-hoc assumption that part of the continuum X-ray emission is occulted, e.g. in a flare partially hidden by the star (Drake et al. 2008); *ii)* in the Class I protostar Elias 29 in the ρ Oph Star Forming Region (SFR) the line was observed during the quiescent state and not during a flare (Giardino et al. 2007), suggesting a production mechanism unrelated to the central star X-ray luminosity and possibly originating from collisional excitation by non-thermal electrons in star-disk loops. In this latter scenario, if the observational viewpoint is favourable, a rotational modulation of the Fe K alpha 6.4 keV line strength could be expected. High-quality time-resolved (on few thousands of seconds) X-ray spectroscopy is thus needed for a better understanding of the connection between the central continuum X-ray source and the emission line.

If the emission occurs on the disk and is directly related to the occurrence of intense flares, the possibility of measuring time delays, expected to be of 50-100 s (for an inner disk radius of 0.1-0.2 AU), with the reverberation mapping approach (either in the X-ray band alone or by simultaneous X-ray and IR observations), can give a firm handle on the physical process at work, as well as on the geometrical structures of the disks, such as inner disk holes, disk gaps and associated walls, perhaps also clumpy disk structures that might be related to planet formation or disk photo-evaporation (cf Feigelson et al. 2009).

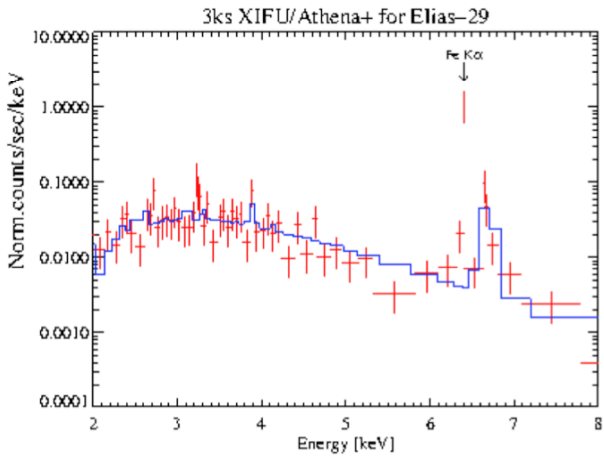

Figure 2: A simulated X-IFU spectrum of the Class I protostar Elias 29 in the range of energies around the Fe K alpha 6.4 keV and the thermal Fe 6.7 keV lines. The model fit to the data does not include, purposely, the Fe K alpha at 6.4 keV. With an exposure time of just 3 ks the presence of a Fe K alpha 6.4 keV line with an equivalent width of ~ 850 eV (as found with a much longer XMM-Newton observation) is easily recognized with a significance level better than 5 sigma.

The X-IFU (or WFI) will enable these time-resolved spectroscopy studies: in Elias 29 the 6.4 keV line can be detected (at > 5 sigma) in only 3 ks (cf. Fig.2) at much higher significance than in the existing XMM-Newton data. With typical rotation periods for YSOs of ≥ 1 d, their rotational cycle can be well sampled with time-resolved spectroscopy. Given the extent of the *Athena+* field of view, the Fe Kα line can be simultaneously detected from several YSOs belonging to the same SFR. Moreover the X-IFU, with its more than 200 times improved "effective area x spectral resolution" (for E > 0.6 keV), will allow a detailed characterization of the X-ray irradiation of YSO circumstellar disks in all nearby SFRs. In general, the *Athena+* collecting area around and above 6 keV will allow us to establish the intensity of deeply penetrating hard (> 7 keV) X-rays needed to evaluate turbulence and the disk "dead" zone.

## 3. EXTREME STELLAR FLARES

Magnetic reconnection events are highly dynamic and involve rapid changes in X-ray-emitting plasma composition, temperature, and bulk velocities. While changes in temperature can be measured from time-resolved low-resolution spectra, making progress in understanding the effects of magnetic reconnection requires constraints on density, elemental abundance, scale length, and velocity of flaring plasma, which are beyond the capability of presently available high-resolution spectrometers. The bulk velocities of the impulsive mass motions which take place early in stellar flares are currently unconstrained; however, based on the solar case, the time scale of those motions could be as short as 10-1000 seconds. The velocities can be derived from spectral line shifts during the short time in which the bulk motion occurs, which is when the X-ray flare is still in the early, rising and therefore faint, phase; averaging in time will smear out the signal and produce ambiguous results. Studying this phenomenon thus requires high collecting area.

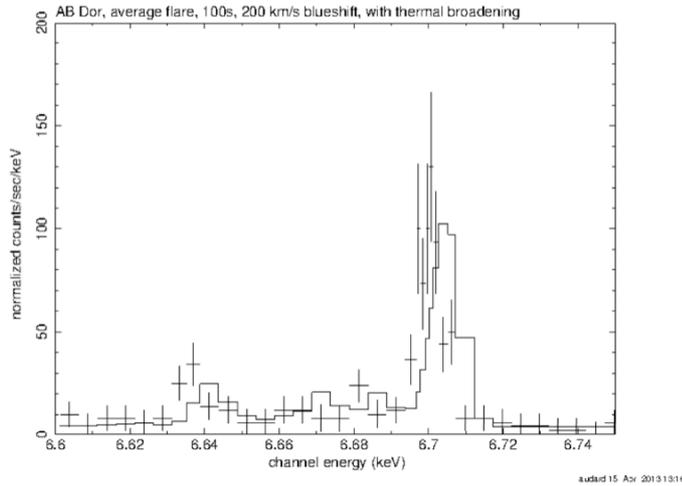

Figure 3: The predicted emission associated with an average flare seen on the nearby dKe star AB Dor (fX ~ 3-7 $10^{-11}$ erg cm$^{-2}$ s$^{-1}$) in 100 s integration time. Such a short integration time will allow us to probe the bulk motion of plasma in the early flare phase. The simulated spectrum assumes a bulk motion of flare-related plasma of 200 km s$^{-1}$ that can be recognized by the resulting Doppler shift of the Fe 6.7 keV line. This is reinforced by the fact that we expect a bulk velocity up to 2 times higher.

The velocities associated with blue-shifts, red-shifts and turbulence of chromospheric evaporation in the early flare phases are typically of the order of 100 – 400 km s$^{-1}$. Such values are accessible with the X-IFU in the more intense stellar sources (f$_X$ ~ $10^{-11}$–$10^{-12}$ erg cm$^{-2}$ s$^{-1}$); particularly interesting are the large, energetic flares seen on stars different from the Sun, such as dMe, T-Tauri stars and protostars with disks, very active zero-age main-sequence stars during the time of planetary system formation, etc..

Those flares should be characterized by very/extremely hot plasma; hence the Fe 6.7 keV line – where the X-IFU resolution will be ~ 3000 – will provide a unique insight on the bulk motion of the plasma (as shown in Fig. 3 for the case of an average flare on AB Dor). Particularly intriguing are the huge flares (peak L$_X$~$10^{32}$ erg s$^{-1}$, peak T > 200MK) seen in CTTSs and occurring in big magnetic structures that, although lacking firm observational evidence, are currently thought to connect the disk (at the co-rotation radius) and stellar surface with sizes > R$_{STAR}$ (e.g. Favata et al. 2005). In such a scenario, the flare-released energy may evaporate the circumstellar disk material at the magnetic loop footpoint, and the bulk motion (cf. Orlando et al. 2011) of the resulting plasma – including the rotational velocity imprint – could be revealed thanks to the X-IFU collecting area and spectral resolution.

## 4. PROBING NON-STANDARD DYNAMOS AND EMISSION MECHANISMS IN ULTRA-COOL DWARFS

Ultra-Cool Dwarfs (UCDs) are objects at the bottom of the stellar main-sequence (spectral type M7 and later) and, depending on the age, can be either stars or brown dwarfs. At young ages these spectral types correspond to substellar masses, i.e. all young UCDs are brown dwarfs (BDs). XMM-Newton and Chandra have established young BDs in star forming regions as ubiquitous X-ray emitters. However, the X-ray luminosities of evolved field BDs are yet unknown and only one L-type brown dwarf has been detected in X-rays to date (Audard et al. 2007). As a result, the nature, the efficiency, and the evolution of the dynamo – of which the coronal emission is a proxy – in substellar objects are still poorly known. Investigations of the occurrence of other X-ray emission mechanisms known to be at work in CTTSs (e.g. accretion shocks, cf. Sect. 2), or in Solar System planets (cf. Branduardi-Raymont, Sciortino, et al. 2013, *Athena+* supporting paper) have been limited by low count statistics.

In solar-like low-mass stars the ratio between X-ray and bolometric luminosity (L$_X$/L$_{Bol}$), which is an indicator of the strength of magnetic activity and of the underlying dynamo action, spans the range $10^{-3...-6}$ with some scatter beyond these limits. In the BD regime, the L$_X$ – L$_{Bol}$ relation is dominated by upper limits, so current data cannot provide a firm assessment of the size of the drop of the X-ray efficiency with respect to low-mass stars and its dependence on fundamental parameters such as effective temperature or rotation. In *young* BDs, possible causes for a decrease of the L$_X$/L$_{Bol}$ level are the transition from a solar-like to a convective dynamo, centrifugal disruption of the corona due to smaller corotation radius or suppression of X-ray emission by accretion (e.g. Jardine et al. 2006; Flaccomio et al. 2003). Among the known young BD population in star forming regions only those with low extinction can be detected using present-day instrumentation at its very limit (cf. the 850 ks long COUP, Preibisch et al. 2005). With *Athena+* we will reach a limiting flux f$_X$ ~ 5 $10^{-17}$ erg cm$^{-2}$ s$^{-1}$ in 100 ks, and detect young BDs in Orion (450 pc) down to L$_X$=$10^{27.2}$ erg s$^{-1}$ even if absorbed by A$_V$ = 10 mag and about twice deeper (or faster) for A$_V$ = 5 mag. With *Athena+*, in the nearby

SFRs (d ~ 150 pc), the vast majority of the upper limits (Grosso et al. 2007) will be turned into detections and the possible drop of $L_X$ vs. $L_{Bol}$ in the substellar regime (see Stelzer et al. 2012a) be constrained.

For the *evolved* UCDs in the field, the chromospheric Hα emission drops steeply beyond a spectral type of about M7. Evidence for an analogous decay of the X-ray emission is based on less than a handful of upper limits for BDs with ages > 100 Myr (Stelzer et al. 2006). A decline of chromospheric and coronal emission could be due to the decrease of effective temperature with time for substellar objects and the related inefficiency of coupling between the increasingly neutral matter and the magnetic field (Mohanty et al. 2002). However, strong radio emission is detected in ~5-10% of the UCDs (Antonova et al. 2012) and the characteristics of their radio emission undergoes a dramatic change: short radio pulses and circular polarization show up that have been ascribed to electron cyclotron maser instability (Hallinan et al. 2008) a mechanism which is also responsible for the aurorae of the solar system planets (Zarka 1998). A critical question is, therefore, to what extent dynamos operating in UCDs produce solar-like coronal magnetic fields, or whether they generate planet-like magnetospheres with completely different physics of magnetic heating, wind mass loss, etc. While a small number of UCDs has been detected in X-rays, the radio-bursting sources seem to be X-ray quiet or very weak (Stelzer et al. 2012b). From the existing observational material, $L_X \leq 10^{25}$ erg s$^{-1}$ must be expected for evolved UCDs. With *Athena+* the detection of such objects is feasible in just ~ 30 ks at a distance of 25 pc. According to the census of nearby stars within 20 pc (Reid et al. 2008), ~ 100 L dwarfs will be accessible to *Athena+* including the third closest neighbour of the Sun, an L dwarf at 2pc (Luhman et al. 2013), and many late-M dwarfs. Given the low count statistics, very few X-ray spectra of BDs outside SFRs, with ages ≥ 10 Myr, are currently available (Rutledge et al. 2000; Tsuboi et al. 2003; Stelzer 2004).

Current X-ray instruments provide poor constraints on the temperature of X-ray emitting plasma in UCDs and high-resolution spectroscopy giving access to the coronal densities is unfeasible. The importance of spectral analysis is evident from the recent Chandra detection of a young BD (Stelzer et al. 2010) that has shown unusually high X-ray luminosity ($L_X \sim 10^{29.7}$ erg s$^{-1}$) combined with unexpectedly soft temperature (kT ~ 0.23 keV). These properties are remarkably similar to those of TW Hya, the prototype of an accreting T Tauri star, whose X-ray emission is dominated by the contribution from accretion shocks with only a minor coronal component.

## 5. MASSIVE STARS: MASS-LOSS, STELLAR WINDS AND MAGNETIC FIELDS

Owing to their powerful stellar winds, which combine large mass-loss rates and wind velocities of several thousand km s$^{-1}$, massive stars are key players in feedback processes on local scales within individual galaxies, both in the nearby and more distant Universe. High-quality multi-wavelength observations of massive stars have shown that the classical picture of a smooth, homogeneous stellar wind is far too simplistic. Real stellar winds feature small and large-scale structures which impact on the conventional diagnostics of mass-loss rates such as Hα emission, UV P-Cygni profiles and free-free radio emission. As a result, discrepancies by one or two orders of magnitude on the mass-loss rates are not uncommon (Fullerton et al. 2006, Oskinova et al. 2007). Solving these issues is fundamental for a proper understanding of massive star evolution, which is heavily dependent on the mass-loss processes, and for a quantitative assessment of massive stars' feedback.

**Probing the dynamics of stellar winds.** Small-scale structures in stellar winds are produced by the same dynamical instabilities of radiatively-driven flows that generate the soft X-ray emission of most single massive stars. As the X-ray emission arises inside the winds, structures impact the X-ray line profiles (Oskinova et al. 2006). High-resolution X-ray spectroscopy thus provides a powerful probe of the mass-loss rates, the degree of wind inhomogeneity and the shape of the structures (Hervé et al. 2013). Chandra and XMM-Newton have paved the way in that direction, but the current sample (only about a dozen objects) of high-quality high-resolution X-ray spectra of massive stars is much too small to achieve a general understanding of the structure of stellar winds. Further progress requires an increase in sensitivity, in conjunction with high spectral resolution. With its unique capabilities, *Athena+* X-IFU holds great potential to provide major breakthroughs in this field, notably by extending measurements beyond the few brightest objects, significantly enlarging the sample of high-resolution spectra and by opening up the field of time-resolved high-resolution spectroscopy.

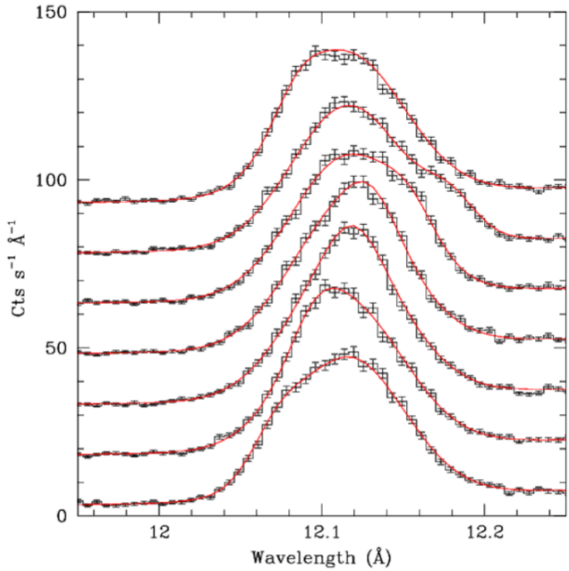

Figure 4: Simulated sequence of seven 1 hour (3.6 ks) *Athena+* X-IFU exposures of the Ne X Lyα line in the spectrum of ζ Pup, assuming a large scale structure produces variations with an amplitude of 10% as observed in XMM-Newton broad-band data.

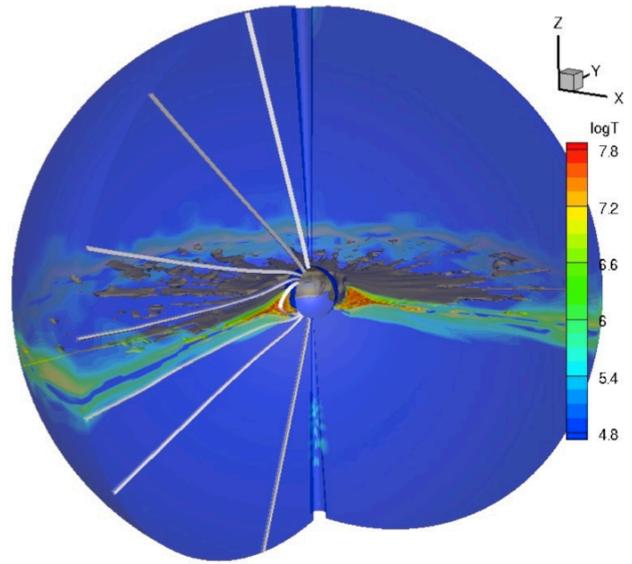

Figure 5: Simulated 3D distribution of the logarithm of temperature in the magnetically confined wind of θ$^1$ Ori C (ud-Doula et al. 2013). The white lines illustrate the magnetic field lines at various magnetic latitudes.

Large-scale structures in stellar winds, co-rotating with the star, are expected to produce X-ray line profile variability on the time scales of the stellar rotation period (in most cases a few days). Existing X-ray facilities provide hints of line profile variations at the 10% level, but lack the sensitivity to collect high-quality data within short enough exposure times (Nazé et al. 2013). *Athena+* X-IFU offers for the very first time the opportunity to properly sample the time scales of these variations with good statistics for the X-ray brightest O-stars (see Fig. 4).

Large scale structures are most likely produced by magnetic fields, leading in the most extreme cases to so-called magnetically confined wind structures (MCWS, Babel & Montmerle 1997, ud-Doula & Owocki 2002). Magnetic fields of up to kG strength are nowadays commonly detected through spectro-polarimetry (see Walder et al. 2011 and references therein). In massive stars featuring an MCWS, a rather hard X-ray emission arises from the head-on collision between the winds flowing from both hemispheres towards the magnetic equator (see Fig. 5). As a result, this hard emission component displays rotational variations that are directly linked to the properties of the MCWS (strength and orientation of the magnetic field). With the X-IFU it will be possible to study the variability of the relevant X-ray emission lines and to Doppler-map the innermost parts of the MCWS of a sample of magnetic massive stars. This will be of utmost importance for further advances in the theory of the generation of magnetic fields in massive stars, and their inclusion in stellar evolution models.

**Probing the physics of wind-wind interactions.** At least 50% of the O-type stars in our Galaxy are in binary systems (Sana et al. 2012). The head-on collision of stellar winds in such systems can produce a harder and stronger X-ray emission than seen in single massive stars (e.g. Stevens et al. 1992). The profiles of X-ray emission lines formed inside the wind interaction zone, such as the strong Fe 6.7 keV line seen in many interacting wind systems, provide a unique probe of the immediate post-shock conditions in the wind interaction zone, directly reflecting the dynamics of the hottest gas (Henley et al. 2003). The high sensitivity and spectral resolution of *Athena+* X-IFU are needed to monitor the orbital changes of these lines which stem from the changing orientation of the system and, in the case of eccentric systems, also the changing properties of the wind interaction as the separation between the stars varies (see Fig. 6). In the particular case of circular orbit systems, Doppler tomography can be used to map the colliding wind region in velocity space. In combination with hydrodynamical models (e.g. Pittard & Parkin 2010), high-resolution X-ray spectra of interacting wind systems constrain the post-shock densities and temperatures in the shock region, hence providing information on the stellar mass-loss rates and wind speeds which are essentially independent of small-scale structures.

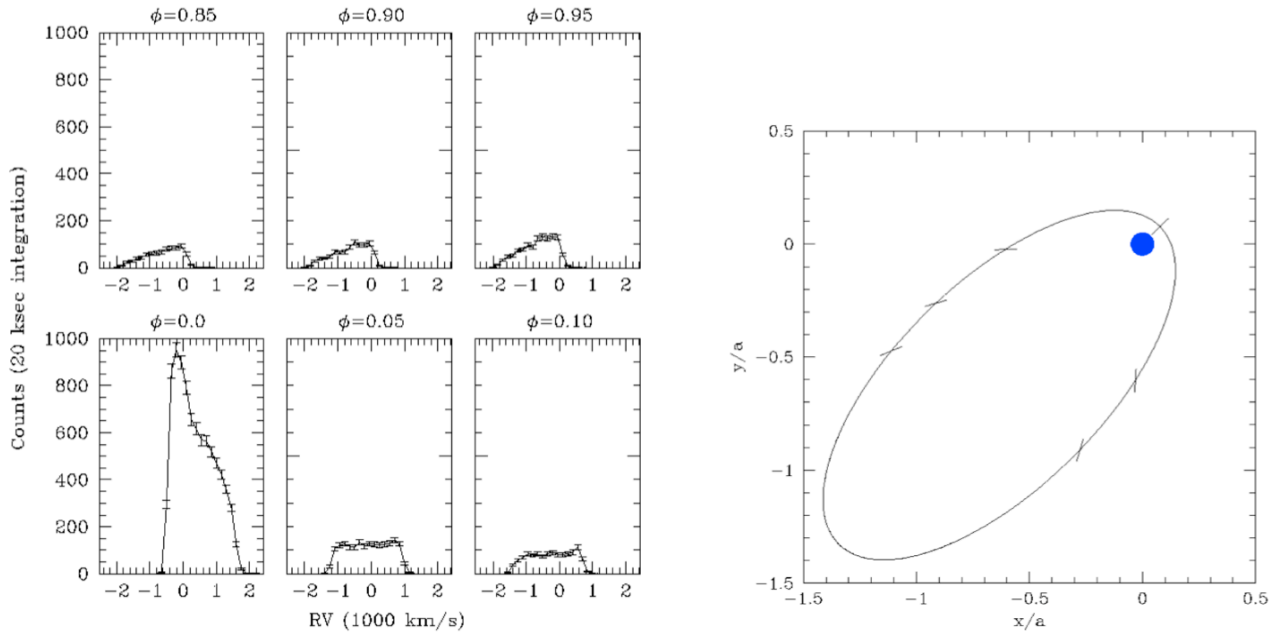

**Figure 6:** *Athena+* X-IFU simulation of the Fe 6.7 keV line emission of the highly eccentric colliding wind binary WR140. [Left] The panels show 20 ks X-IFU simulations of the Fe 6.7 keV line (shown in radial velocity space) at six distinct phases around periastron. The changes of the line intensity reflect the changing orbital separation, whilst the changes in the line profile are mainly due to the changing orientation of the shock cone. [Right] the orbit of WR140. The position of the star with the weaker wind is given by the tick marks at those orbital phases simulated in the left panel. The observer is located at the bottom.

**Probing the extremes of stellar winds.** The majority of the O-type stars follow a scaling relation between X-ray and bolometric luminosity, $\log(L_X/L_{Bol}) \sim -7$. The origin of this relationship has been a mystery for many years, but could stem from mixing of cool and hot material in the turbulent post-shock plasma (Owocki et al. 2013). The situation is less clear-cut for B-type stars which have weaker winds and for Wolf-Rayet stars and Luminous Blue Variables (Nazé et al. 2012) which have much stronger and denser winds. Some of the objects of the latter categories are detected, whilst others are not. Within a 20 ks exposure, *Athena+* WFI will allow us to go an order of magnitude deeper than current upper limits, thereby permitting the detection of the predicted wind emission. This will provide new insight into the physics of the extremes of massive stars' winds.

With a few exceptions, current studies of X-ray emission from massive stars are restricted to objects in our own Galaxy. Owing to its unique combination of effective area and spatial resolution, the *Athena+* WFI will enable us to extend these studies beyond the Galaxy, notably to the Magellanic Clouds, thereby allowing us investigate the impact of metallicity on the physics of stellar winds. Indeed, the winds of massive stars are driven by resonance scattering in metal lines, hence metallicity is expected to play a key role (de Koter 2008). *Athena+* WFI will also allow the study of massive star clusters in star-forming galaxies with unprecedented details, revealing the combined feedback from the clusters, through wind-blown bubbles, galactic fountains and super-shells filled with X-ray plasma.

## 6. PLANETARY NEBULAE, PROBING THE FUTURE OF OUR SUN

Planetary nebulae (PNe) are shells of enriched stellar material, lost during the asymptotic giant branch (AGB) phase of intermediate-mass stars, swept up by subsequent fast winds and ionized by UV radiation from the newly exposed stellar core. Chandra and XMM-Newton have revealed many examples of diffuse X-ray emission from the shocked fast winds in PN interiors ("hot bubbles"). The X-ray spectra of the latter are soft and the implied plasma temperatures, ≤ 3 MK, are much lower than the post-shock temperatures expected given the observed wind velocities of 1000-3000 km s$^{-1}$. High-resolution spectra of the faint, soft, diffuse X-ray emission from PNe are needed to accurately determine their interior plasma abundances and temperatures, and to constrain the wind interaction processes that generate PN hot bubbles. To date, however, only the two X-ray brightest PNe have been observed in high-resolution X-ray spectroscopy: BD+30° 3639 (Fig. 7) and NGC 6543, both requiring exposure times of several hundred ks. The sensitivity and spectral resolution of *Athena+* X-IFU will dramatically increase the feasibility of high S/N high-resolution spectroscopy of diffuse X-ray emission from PNe (Fig. 8), extending the sample to about 30 PNe (Kastner et al. 2012). These spectra will allow us to use resolved spectral lines to assess the roles played by physical processes such as heat conduction, charge exchange, and mass loading in determining the abundances and moderating the temperatures of PN hot bubbles.

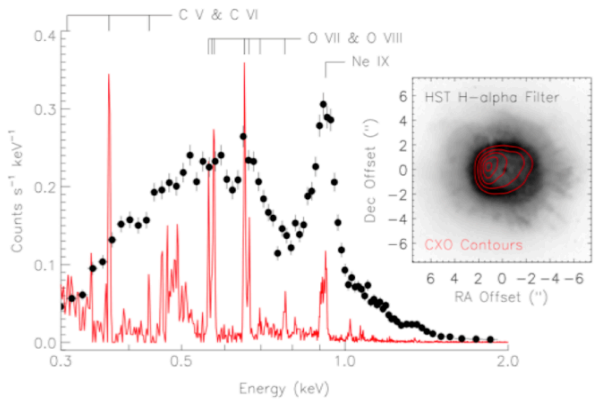
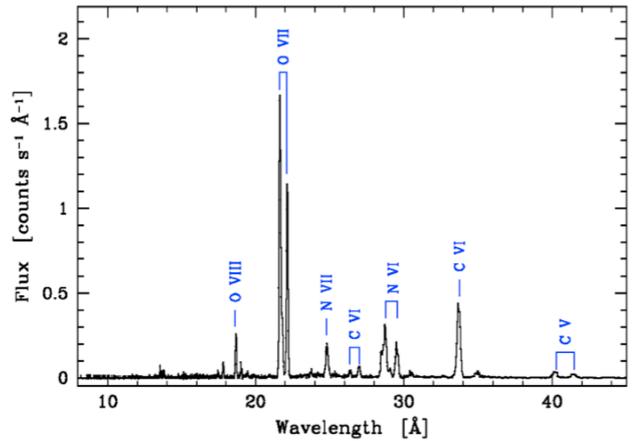

Figure 7: **Chandra X-ray observations of the compact PN BD +30°3639.** The insert shows contours of the X-ray-emitting hot bubble which fits snugly within the dense nebular rim imaged by HST in Hα (greyscale). Strong Ne emission from the bubble is apparent in the ACIS CCD spectrum (black circles). The grating data (red) further reveal that the low-energy end of the spectrum is dominated by C and O emission lines, and constrain its temperature to lie in the range 1.7-2.9 MK (Yu et al. 2009). The forest of lines near 0.5 keV is likely due to $C^{6+}$ ions from the hot bubble penetrating exterior (cooler) nebular gas before recombining (Nordon et al. 2009).

Figure 8: **Simulated 20 ks *Athena+* X-IFU spectrum of the PN NGC 6543 (the Cat's Eye nebula)**, adopting plasma parameters determined from Chandra and XMM data (i.e., kT = 0.1 keV, absorbed flux = $10^{-13}$ erg cm$^{-2}$ s$^{-1}$). The predicted X-IFU background subtracted count rate is 0.976 cts s$^{1}$, for a total net count number of 19,500 ± 150. X-IFU will extend such detailed studies of abundances and physical conditions to many more PNe than presently feasible.

In addition to diffuse X-ray emission from PN wind interactions, about half of the PN central stars (CSPN) emit X-rays (Kastner et al. 2012). Some CSPN show soft X-ray emission as expected from the 100-200 kK photosphere, but the vast majority also exhibit harder X-ray emission, often characterized by 1 to 40 MK thermal plasma. These sources may indicate the presence of binary companions to CSPNe, although such an explanation evidently does not apply to all objects (e.g., the CSPN of the Helix Nebula). *Athena+* can collect high-resolution spectra of the CSPN and allow comparisons with non-local thermodynamic equilibrium model atmospheres or simulations of CSPN accretion and wind shock processes. Only through high-resolution spectra can we diagnose the nature and emission mechanism of the hard X-rays from CSPN. The high sensitivity of *Athena+* will further allow us to obtain light curves for X-ray-emitting CSPNe and, via emission line radial velocity analyses, assess whether binary companions or variable stellar winds are responsible for the X-ray emission.